%% file: main.tex
\documentclass[journal]{IEEEtran}

\usepackage[T1]{fontenc}          
\usepackage{amsmath,amssymb}      
\usepackage{physics}              
\usepackage{graphicx}             
\usepackage{booktabs}             
\usepackage{amsthm}               
\usepackage{cite}                 
\usepackage[hidelinks]{hyperref}  

\newtheorem{assumption}{Assumption}
\newtheorem{proposition}{Proposition}

\begin{document}

\title{A Physics-Grounded QUBO Encoding of Irrigation Scheduling for QAOA}

\author{Alisher~Ortikov and Alisher~Ilhamov%
\thanks{July~14,~2026.}%
\thanks{A. Ortikov is with the School of Computing, KAIST, Daejeon, Republic of Korea (e-mail: mcpeblocker@kaist.ac.kr). (Corresponding author: Alisher Ortikov.)}%
\thanks{A. Ilhamov is with Westminster International University in Tashkent, Tashkent, Uzbekistan (e-mail: alisherilhamov44@gmail.com).}}

\maketitle

\begin{abstract}
Rotational irrigation scheduling in water-scarce Central Asia is a densely
coupled combinatorial problem: soil-moisture memory links each irrigation
decision to all later days within a zone, field adjacency couples zones on
overlapping window days, and rigid canal rotations quantize water delivery
in time. We formulate it as a Quadratic Unconstrained Binary Optimization
(QUBO) by linearizing the root-zone water balance, so the quadratic
crop-stress objective generates the physical couplings as 2-local Ising
interactions with no higher-order terms; only the water-budget constraint
requires an artificial all-to-all penalty, which we certify with an
instance-adaptive weight bound an order of magnitude tighter than generic
prescriptions. Every instance is built from observed data for a cotton
district in Khorezm, Uzbekistan: NASA POWER meteorology, FAO-56
Penman--Monteith evapotranspiration, SoilGrids~2.0 hydraulics, measured
capillary fluxes, and documented canal-rotation windows that enter as
qubit-count reductions. We benchmark four tiers --- exact solvers,
matched-budget heuristics, ideal-statevector quantum approximate
optimization algorithm (QAOA), and noise-model plus IBM
Heron execution --- and add a scaling study on soil instances up to 584
variables. Exact branch-and-bound proves optimality in seconds through 150
variables, and heuristic degradation at fixed evaluation budget is repaired
by scaling the budget, so no classical scalability wall appears at
deployment-relevant sizes, and none is claimed. On hardware, the
informative signal is optimum-sampling enrichment over uniform sampling. We
claim no quantum advantage; we deliver a physically grounded,
data-complete, reproducible encoding of a societally critical scheduling
problem for the quantum-utility era.
\end{abstract}

\begin{IEEEkeywords}
Combinatorial optimization, irrigation scheduling, near-term quantum
devices, quadratic unconstrained binary optimization (QUBO), quantum
approximate optimization algorithm (QAOA), quantum computing.
\end{IEEEkeywords}

\input{sec_introduction}
\input{sec_related_work}
\input{sec_problem_formulation}
\input{sec_methods}
\input{sec_results}

\input{sec_limitations}
\input{sec_conclusion}

\input{app_derivation}
\input{app_experiments}

\input{sec_acknowledgments}

\bibliographystyle{IEEEtran}
\bibliography{refs}


\end{document}

%% file: sec_introduction.tex
\section{Introduction}
\IEEEPARstart{A}{gricultural} sustainability in Central Asia, particularly within Uzbekistan, is experiencing an unprecedented crisis driven by severe water scarcity. The region's agricultural backbone relies almost entirely on intensive irrigation networks fed by two major transboundary rivers: the Amu Darya and the Syr Darya. Decades of over-exploitation, coupled with accelerating climate change and the historic ecological collapse of the Aral Sea basin, have made traditional, rigid water management strategies completely unsustainable \cite{oguz2026socio,micklin2007aral,saidmamatov2023nexus}. Because agriculture accounts for over 83\% of the country's total freshwater consumption, even marginal improvements in water-use efficiency can yield massive macro-economic and ecological relief \cite{fao_aquastat,saidmamatov2023nexus}. At the same time, the economic feasibility of modernized irrigation delivery (including farmer willingness to pay for improved services) remains a practical constraint on deployment \cite{kassie2022willingness}. Transitioning from uniform, calendar-based flood irrigation to data-driven precision agriculture is no longer just an optimization goal --- it is an existential necessity for the region \cite{adb2020uzbekistan,worldbank2025uzbekistan}.

While modern precision agriculture leverages sensor networks and weather forecasts to inform scheduling, optimizing these decisions at scale is a genuinely combinatorial task \cite{abioye2020review}. A realistic irrigation schedule cannot treat fields or days in isolation. Instead, it must account for two coupled physical phenomena:
\begin{itemize}
    \item \emph{Temporal coupling}: soil exhibits ``moisture memory,'' meaning an irrigation event today directly governs the baseline soil moisture, evapotranspiration response, and crop stress profiles of every subsequent day. In the instances constructed here this intra-zone memory is the \emph{dominant} coupling structure.
    \item \emph{Spatial coupling}: adjacent zones interact through lateral moisture movement, runoff, and shared offtake capacity. Under realistic staggered canal rotations, however, adjacent zones can co-irrigate only on the few days where their windows overlap, so spatial terms are sparse compared to the temporal cliques.
\end{itemize}
These physical couplings combine with the operational constraints of Uzbekistan's water infrastructure --- rigid, state-managed canal networks that restrict water availability to specific rotational windows \cite{adb2020uzbekistan}. The result is a densely coupled binary optimization problem whose temporal interaction structure inflates classical branch-and-bound trees and has motivated an extensive irrigation-scheduling optimization literature, spanning simulation-optimization crop models \cite{liu2021irrigation} and evolutionary metaheuristics \cite{zhang2022canal,maier2014metaheuristics}.

We wish to be explicit about what this paper does \emph{not} claim: it does not demonstrate that these instances are classically intractable, and it reports no quantum advantage. Our own classical scaling study (Section~\ref{sec:scaling}) shows that an off-the-shelf mixed-integer quadratic programming (MIQP) solver proves global optimality in seconds far beyond the sizes any near-term quantum device can address, and that the apparent degradation of matched-budget metaheuristics with instance size is an evaluation-budget artifact rather than a hardness wall. The motivation for a quantum formulation is instead structural and forward-looking. The scheduling decision --- whether to activate an irrigation event for a specific zone on a specific day --- is inherently discrete, and by unrolling the linear soil-moisture dynamics, the cross-day and cross-zone physical interactions emerge as pairwise quadratic terms ($x_i x_j$) with no higher-order polynomials. The physical objective therefore maps onto a 2-local cost Hamiltonian executable by the Quantum Approximate Optimization Algorithm (QAOA) \cite{farhi2014qaoa} without Trotter error, and the encoding, its certified penalty bound, and its measured hardware behavior become reusable assets for the quantum-utility era --- when problem sizes, recourse structure, or tighter re-planning loops may change the classical picture.

In this paper, we bridge the gap between theoretical quantum optimization and a fully data-grounded agricultural problem. Our explicit contributions are fivefold:
\begin{itemize}
    \item \textbf{A physics-grounded QUBO formulation.} A linear ``bucket'' soil-moisture model whose quadratic crop-stress objective \emph{derives} the temporal and spatial couplings from hydrological principles, together with a certified, instance-adaptive budget-penalty bound (Proposition~\ref{prop:penalty}) that is an order of magnitude tighter than generic prescriptions.
    \item \textbf{Real-data instance construction.} Every model coefficient is computed from observed data for a cotton district in Khorezm, Uzbekistan: NASA POWER (MERRA-2) daily meteorology, FAO-56 Penman--Monteith evapotranspiration, SoilGrids~2.0 soil hydraulics, and field-measured shallow-groundwater capillary fluxes \cite{nasapower,allen1998fao56,soilgrids,forkutsa2009modeling}.
    \item \textbf{Central Asian operational grounding.} The rigid rotational canal schedules of Uzbek Water Consumers Associations enter as hard variable eliminations --- reducing qubit count rather --- and shallow-groundwater dynamics unique to the lower Amu Darya enter the water balance \cite{adb2020uzbekistan,veldwisch2008cotton}.
    \item \textbf{A transparent benchmark with an acknowledged classical frontier.} A four-tier evaluation (exact enumeration and exact MIQP, matched-budget classical heuristics, ideal-statevector QAOA, and noise-model plus IBM Heron hardware execution), extended by a classical scaling ladder to $n=584$ variables on real soil data that candidly locates \emph{no} classical scalability wall at deployment-relevant sizes (Section~\ref{sec:scaling}).
    \item \textbf{A constraint-preserving XY-mixer variant.} A new formulation (Section~\ref{sec:xy}) that restricts evolution to the budget-feasible manifold via an XY ring mixer and Dicke initialization, eliminating the budget penalty and its slack qubits entirely; we derive it, verify it against the full circuit, and quantify on hardware both where it helps (per-shot optimum sampling, gate count at fixed small width) and where its state-preparation overhead dominates (larger widths, depth) --- both outcomes reported as measured.
\end{itemize}

%% file: sec_related_work.tex
\section{Related Work}

\subsection{Classical Irrigation Management Systems}
The transition from calendar-based irrigation to precision agriculture has been largely driven by the integration of IoT sensor networks and classical computational optimization \cite{abioye2020review}. In contemporary agricultural management systems, crop water requirements are typically forecasted using established meteorological baselines, such as the FAO Penman--Monteith method, and subsequently fed into decision-support systems \cite{allen1998fao56,fao_cropwat}. 

To optimize these scheduling decisions, classical frameworks rely heavily on metaheuristic algorithms, primarily Genetic Algorithms (GA), Particle Swarm Optimization (PSO), and Ant Colony Optimization (ACO). While these classical solvers perform adequately for isolated fields with continuous, on-demand water access, they encounter critical limitations when applied to the operational realities of Central Asian hydrology. Specifically, when irrigation schedules must respect rigid, multi-day canal rotational windows and account for the continuous ``moisture memory'' of the soil, the decision variables become densely coupled across time and space \cite{adb2020uzbekistan}.

This spatiotemporal coupling makes the scheduling task a densely
interacting binary optimization problem: the configuration space grows as
$2^{ZD}$ and the moisture-memory terms couple every pair of decision days
within a zone. The metaheuristic literature reports sensitivity to
premature convergence in flat penalty landscapes and substantial tuning
effort for constrained instances \cite{maier2014metaheuristics,nguyen2017irrigation,zhang2022canal}.
We caution, however, against reading such reports as evidence of
intractability at operational sizes: our own scaling experiments
(Section~\ref{sec:scaling}) show that modern branch-and-bound MIQP solvers
prove global optimality in seconds for instances well beyond current
quantum-hardware reach, and that metaheuristic degradation under a fixed
evaluation budget disappears when the budget scales with instance size.
The open question for this problem family is therefore not present-day
classical hardness but whether a physically faithful encoding can ride
improving quantum hardware into regimes --- stochastic recourse
formulations, tighter re-planning loops, or much larger networks --- where
the classical toolchain becomes the binding constraint.

\subsection{Quantum Optimization in Agriculture}
To address the intractable scaling of dense combinatorial problems, the Quantum Approximate Optimization Algorithm (QAOA) has emerged as a leading hybrid-variational heuristic designed for Noisy Intermediate-Scale Quantum (NISQ) devices \cite{farhi2014qaoa}. By natively mapping discrete decision variables to qubits and encoding constraints as interacting terms in a cost Hamiltonian, QAOA offers a fundamentally different computational approach to traversing complex penalty landscapes.

While the theoretical foundations of mapping NP-hard problems to Quadratic Unconstrained Binary Optimization (QUBO) frameworks are well-established in the literature, the intersection of QAOA and precision agriculture remains an underexplored niche. Recent systematic reviews of quantum utility in agriculture have primarily focused on continuous-variable tasks, such as Variational Quantum Eigensolvers (VQE) for chemical fertilizer design, or Quantum Machine Learning (QML) for crop yield prediction.

The limited existing work on quantum resource allocation in agriculture typically relies on highly abstracted, uncoupled benchmark graphs that ignore the physics of the soil. Specifically, prior applications lack the integration of temporal soil-moisture memory and spatial drainage --- the couplings that give the problem its dense pairwise structure. This paper fills that explicit gap by introducing a physics-grounded QUBO formulation that naturally inherits cross-day coupling from linear hydrological dynamics, providing a rigorously benchmarked QAOA implementation --- with an honest classical baseline --- tailored to the strict operational constraints of Central Asian irrigation networks.

%% file: sec_problem_formulation.tex
\section{Problem Formulation}
\label{sec:formulation}

We formalize rotational irrigation scheduling as a Quadratic Unconstrained
Binary Optimization (QUBO) problem. The guiding principle is that the
\emph{physical} couplings of the irrigated field --- soil-moisture memory
within each zone and same-day interaction between adjacent zones --- must
emerge from the hydrology rather than be imposed by hand. Two modeling
concessions are made for quantum compatibility: (i) the soil-water balance
is linearized (Assumption~\ref{as:linear}), which keeps the resulting
Hamiltonian 2-local, and (ii) crop water stress is penalized by a
\emph{symmetric} quadratic deviation from the moisture target, whereas
real crop response is asymmetric --- operational FAO-56 practice applies a
piecewise stress coefficient $K_s$ that activates only under deficit
\cite{allen1998fao56}. We return to both concessions in
Section~\ref{sec:limitations}. One further coupling is \emph{not} physical:
the water-budget constraint enters as a quadratic penalty that couples
every decision variable to every other and to the slack register
(Section~\ref{sec:qubo-total}), and this artificial all-to-all term ---
not the hydrology --- dominates the hardware routing cost
(Section~\ref{sec:results}). Full derivations are deferred to
Appendix~\ref{app:derivation}.

\subsection{Decision Variables and Operational Constraints}
Consider $Z$ agricultural zones (fields served by a common distributary
canal) scheduled over a horizon of $D$ days, chosen to span one rotation
cycle of the Water Consumers Association (WCA). The binary decision variable
\begin{equation}
    x_{z,d}\in\{0,1\},\qquad z\in\{1,\dots,Z\},\; d\in\{1,\dots,D\},
\end{equation}
indicates whether zone $z$ receives an irrigation event of fixed net
application depth $a$ (mm) on day $d$. A uniform dose is standard practice
in Uzbek furrow irrigation, where per-event application norms are fixed by
hydromodule zoning regulations \cite{adb2020uzbekistan,forkutsa2009modeling};
the non-uniform-dose generalization is given in
Appendix~\ref{app:derivation}.

Rotational water delivery restricts each zone to its canal window
$\mathcal{W}_z\subseteq\{1,\dots,D\}$; outside it the offtake is physically
closed. We enforce this \emph{hard} constraint by variable elimination in
classical pre-processing, fixing $x_{z,d}=0$ for $d\notin\mathcal{W}_z$
rather than penalizing it. Let
$\mathcal{W}=\{(z,d):d\in\mathcal{W}_z\}$ denote the surviving decision
index set; only $|\mathcal{W}|\le ZD$ logical variables reach the quantum
processor.

\subsection{Linear Soil-Water Dynamics}
Let $M_{z,d}$ (mm) be the plant-available water stored in the root zone of
zone $z$ at the end of day $d$, measured on the available-water scale:
$M=0$ at permanent wilting point and $M=\mathrm{TAW}_z$ (total available
water) at field capacity, with
$\mathrm{TAW}_z=(\theta^{\mathrm{FC}}_z-\theta^{\mathrm{WP}}_z)\,Z_r$
for volumetric water contents $\theta$ and rooting depth $Z_r$
\cite{allen1998fao56}.

\begin{assumption}[Linear bucket dynamics]\label{as:linear}
Within the operating regime maintained by the scheduler, the root-zone
water balance evolves linearly,
\begin{equation}
    M_{z,d} = M_{z,d-1} + a\,x_{z,d} + P^{\mathrm{eff}}_{z,d}
              + G_{z,d} - \mathrm{ET}^{c}_{z,d},
    \label{eq:bucket}
\end{equation}
without saturation clipping, where $P^{\mathrm{eff}}_{z,d}$ is effective
rainfall, $G_{z,d}$ the upward capillary flux from the shallow groundwater
table characteristic of the lower Amu Darya
\cite{forkutsa2009modeling}, and $\mathrm{ET}^{c}_{z,d}=K_c(d)\,
\mathrm{ET}^{0}_{z,d}$ the crop evapotranspiration
\cite{allen1998fao56}.
\end{assumption}

The omission of clipping at $M=\mathrm{TAW}_z$ (deep percolation) and $M=0$
is deliberate: excursions above field capacity represent water losses and
excursions toward wilting represent crop damage, and both are
\emph{quadratically penalized} by the objective below, so the optimizer is
driven away from the regime where clipping would bind. We revisit the
residual model error in Section~\ref{sec:limitations}. Because
$P^{\mathrm{eff}}$, $G$, and $\mathrm{ET}^{c}$ are computed classically from
observed data, unrolling \eqref{eq:bucket} gives the closed form
\begin{align}
    M_{z,d} &= M_{z,0} + a\!\!\sum_{d'\le d,\,(z,d')\in\mathcal{W}}\!\!x_{z,d'}
              \;+\; \Phi_{z,d},
    \label{eq:closedform}\\
    \Phi_{z,d} &:= \sum_{d'=1}^{d}\big(P^{\mathrm{eff}}_{z,d'}+G_{z,d'}
                  -\mathrm{ET}^{c}_{z,d'}\big),\nonumber
\end{align}
which is \emph{affine} in the decision vector $x$ with a deterministic,
data-driven forcing term $\Phi_{z,d}$. This affinity is the crux of the
formulation: any quadratic function of $M$ remains quadratic in $x$.

\subsection{Objective}
The scheduler balances water consumption against crop water stress:
\begin{equation}
    F(x) \;=\; \underbrace{c_w\, a \!\!\sum_{(z,d)\in\mathcal{W}}\!\! x_{z,d}}_{H_{\mathrm{water}}}
    \;+\;
    \underbrace{\sum_{z=1}^{Z}\sum_{d=1}^{D} w_{z,d}
    \big(M_{z,d}(x) - T_{z,d}\big)^{2}}_{H_{\mathrm{stress}}},
    \label{eq:objective}
\end{equation}
where $c_w$ prices delivered water (pumping, labor, and service fees),
$T_{z,d}$ is the stage-dependent moisture target
$T_{z,d}=\big(1-\tfrac{\rho}{2}\big)\mathrm{TAW}_z$ with soil-water depletion
fraction $\rho$ \cite{allen1998fao56}, and $w_{z,d}$ weights stress by the
yield-response factor $K_y$ of the current phenological stage
\cite{doorenbos1979yield}. The symmetric square in
$H_{\mathrm{stress}}$ penalizes over-wet and under-wet excursions
equally; this is the quadratic-compatibility concession (ii) above, since
agronomic stress functions are asymmetric about the target
\cite{allen1998fao56}. Substituting \eqref{eq:closedform} and
writing $\delta_{z,d}:=M_{z,0}+\Phi_{z,d}-T_{z,d}$ for the schedule-free
moisture deficit, the stress term expands (using $x^2=x$) into
\begin{align}
    H_{\mathrm{stress}} ={}& \sum_{(z,d')\in\mathcal{W}}
    \big(a^{2}S_{z,d'} + 2a\,V_{z,d'}\big)\,x_{z,d'} \nonumber\\
    &+ 2a^{2}\!\!\sum_{\substack{(z,d'),(z,d'')\in\mathcal{W}\\ d'<d''}}\!\!
    S_{z,d''}\; x_{z,d'}\,x_{z,d''} \;+\; \mathrm{const},
    \label{eq:stress-qubo}
\end{align}
with cumulative-weight kernels
\begin{equation}
    S_{z,d'} := \sum_{d= d'}^{D} w_{z,d},
    \qquad
    V_{z,d'} := \sum_{d= d'}^{D} w_{z,d}\,\delta_{z,d}.
\end{equation}
Equation~\eqref{eq:stress-qubo} makes the central physical claim of this
paper explicit: soil moisture memory generates an all-to-all pairwise
coupling between the irrigation days of each zone, with coefficient
$2a^{2}S_{z,d''}$ decaying toward the end of the horizon --- exactly the
dense temporal structure that inflates classical branch-and-bound trees,
yet still strictly 2-local.

\subsection{Constraints as Penalty Terms}
\textbf{Water budget.} Scarcity limits the rotation cycle to at most $K$
irrigation events ($aK$ mm of delivered water):
$\sum_{(z,d)\in\mathcal{W}} x_{z,d}\le K$. We convert this inequality to an
equality via a binary-encoded slack variable
$s(y)=\sum_{k=0}^{m-1} c_k\,y_k$ with $y_k\in\{0,1\}$,
$m=\lceil\log_2(K{+}1)\rceil$, and bounded coefficients
$c_k=2^{k}$ for $k<m{-}1$, $c_{m-1}=K-2^{m-1}+1$, so that
$s(y)$ ranges exactly over $\{0,\dots,K\}$ \cite{lucas2014ising}:
\begin{equation}
    H_{\mathrm{budget}} = \Big(\sum_{(z,d)\in\mathcal{W}} x_{z,d}
    + s(y) - K\Big)^{2}.
    \label{eq:budget}
\end{equation}

\textbf{Spatial coupling.} Simultaneous irrigation of hydrologically
adjacent zones $(z,z')\in\mathcal{E}$ (sharing a lateral boundary or an
offtake of limited capacity) causes runoff, waterlogging, and flow
competition; we penalize same-day co-irrigation over the field adjacency
graph $\mathcal{E}$:
\begin{equation}
    H_{\mathrm{spatial}} = \sum_{d=1}^{D}\sum_{(z,z')\in\mathcal{E}}
    x_{z,d}\,x_{z',d}.
\end{equation}

\textbf{Timing.} Consecutive-day irrigation of the same zone is penalized
(equipment cooldown, labor logistics, and infiltration lag):
\begin{equation}
    H_{\mathrm{timing}} = \sum_{z=1}^{Z}\sum_{d=1}^{D-1}
    x_{z,d}\,x_{z,d+1}.
\end{equation}
$H_{\mathrm{spatial}}$ and $H_{\mathrm{timing}}$ encode operational
\emph{costs}, not feasibility conditions; their weights are physical
preferences rather than constraint enforcers.

\subsection{The Complete QUBO and Penalty Sufficiency}
\label{sec:qubo-total}
Collecting terms, the total cost function over
$n=|\mathcal{W}|+m$ binary variables reads
\begin{equation}
\begin{aligned}
    H(x,y) &= H_{\mathrm{water}} + H_{\mathrm{stress}} + \lambda_B\, H_{\mathrm{budget}} + \lambda_S\, H_{\mathrm{spatial}} \\
           &\quad + \lambda_T\, H_{\mathrm{timing}} .
\end{aligned}
\label{eq:total}
\end{equation}

\begin{proposition}[Budget penalty sufficiency]\label{prop:penalty}
Write $H_{\mathrm{obj}}:=H_{\mathrm{water}}+H_{\mathrm{stress}}
+\lambda_S H_{\mathrm{spatial}}+\lambda_T H_{\mathrm{timing}}
=\sum_i \ell_i x_i + \sum_{i<j} q_{ij} x_i x_j + \mathrm{const}$ in QUBO
form. All pairwise couplings of $H_{\mathrm{obj}}$ are nonnegative by
construction ($q_{ij}\in\{2a^2 S_{z,d''},\,\lambda_S,\,\lambda_T\}\ge 0$).
If
\begin{equation}
    \lambda_B \;>\; \max\Big(0,\; -\min_{i\in\mathcal{W}} \ell_i\Big),
\end{equation}
then every global minimizer of $H$ satisfies the budget constraint.
\end{proposition}
\noindent The proof (Appendix~\ref{app:derivation}) removes excess
irrigation events one at a time: each removal raises the objective by at
most $\max(0,-\min_i\ell_i)$ because couplings are nonnegative, while the
integer-valued penalty drops by at least $\lambda_B$. This per-event bound
is far tighter than the standard objective-range prescription
$\lambda_B>\max H_{\mathrm{obj}}-\min H_{\mathrm{obj}}$
\cite{lucas2014ising}, which for our instances over-penalizes by more than
an order of magnitude. After the normalization required to fix the variational parameter scale, an inflated $\lambda_B$ compresses the agronomic signal into a narrow band of the
spectrum, degrading both QAOA trainability and noise resilience. We set
$\lambda_B$ to $1.1\times$ the certified bound.

\subsection{Ising Mapping and Quantum Resource Count}
Substituting $x_i=(1-Z_i)/2$ for each of the $n$ binary variables maps
\eqref{eq:total} to a 2-local Ising Hamiltonian
\begin{equation}
    H_C = \sum_{i=1}^{n} h_i\, Z_i
    + \sum_{i<j} J_{ij}\, Z_i Z_j + E_0,
    \label{eq:ising}
\end{equation}
with coefficients given in closed form in Appendix~\ref{app:derivation}.
The QAOA phase-separation layer therefore compiles to one $R_{ZZ}$ gate per
nonzero $J_{ij}$ and one $R_Z$ per qubit. The interaction graph has three
structural components: (i) a clique over each zone's open window
(temporal moisture memory), (ii) same-day edges between adjacent zones
(spatial coupling), and (iii) a global coupling of all decision variables
to the slack register (budget). Canal-window elimination directly removes
qubits, so the operational rigidity of Central Asian water delivery ---
ordinarily a complication --- becomes a resource-count \emph{advantage} on
NISQ hardware.

%% file: sec_methods.tex
\section{Methods}
\label{sec:methods}

We evaluate the formulation of Section~\ref{sec:formulation} on problem
instances built entirely from observed field data, benchmarked across four
tiers of solvers: exact enumeration, classical heuristics, idealized QAOA,
and noisy/hardware QAOA.

\subsection{Real-Data Problem Instances}
\label{sec:instances}
All instances describe three adjacent cotton fields along a distributary of
the Shavat canal system near Urgench, Khorezm province, Uzbekistan
(41.55--41.59$^\circ$N, 60.58--60.66$^\circ$E) during the 2024 growing
season (sowing 15 April). Daily meteorology (2-m temperature extrema,
relative humidity, wind speed, all-sky shortwave irradiance, precipitation)
is taken from the NASA POWER agroclimatology archive (MERRA-2 assimilation)
\cite{nasapower}; reference evapotranspiration is computed with the FAO-56
Penman--Monteith equation \cite{allen1998fao56}, giving a season mean
$\mathrm{ET}^0=7.6$ mm\,day$^{-1}$ and 93 mm of season precipitation ---
the extreme aridity that makes Khorezm scheduling unforgiving. Zone-level
soil water constants derive from SoilGrids 2.0 \cite{soilgrids} at the
three field locations
($\mathrm{TAW}\in\{118,106,95\}$ mm over a 1.0 m root zone), and the
capillary-rise flux uses the mid-season mean measured in Khorezm cotton
fields with shallow water tables \cite{forkutsa2009modeling}. Full
parameter provenance is tabulated in Appendix~\ref{app:instance}.

The scheduling horizon covers peak crop water demand (July, flowering
stage) with staggered 3-day canal windows per zone, overlapping by one day,
following documented WCA rotation practice
\cite{adb2020uzbekistan,veldwisch2008cotton}. Three tiers are used:
\emph{small} ($D=7$, one rotation, $K=2$; $9$ decision $+\,2$ slack $=11$
qubits), \emph{medium} ($D=14$, two rotations, $K=4$; $18+3=21$ qubits),
and \emph{large} ($D=28$, four rotations, $K=8$; $36+4=40$ variables,
classical solvers only). For the classical scaling study
(Section~\ref{sec:scaling}) this ladder is extended to
$n\in\{77,150,295,584\}$ by adding zones along the Urgench--Khiva
irrigation corridor and further rotation cycles: SoilGrids~2.0 is sampled
at 24 field locations, zones are grouped into parallel
three-zone distributary branches with the same staggered windows, and
zones beyond the original three reuse the zone-1 meteorological forcing
--- justified because the three base zones already share a single MERRA-2
grid cell, i.e.\ zone heterogeneity at this spatial scale comes from soil,
not weather (Section~\ref{sec:limitations}).

\subsection{Classical Baselines}
\label{sec:classical-methods}
For $n\le 21$ the global optimum is obtained by vectorized exhaustive
enumeration, which also certifies (per Proposition~\ref{prop:penalty}) that
the penalized optimum is budget-feasible. Heuristic baselines are:
\begin{enumerate}
    \item \textbf{Greedy agronomic rule}: the FAO irrigation-trigger
    heuristic used in operational practice --- each day, irrigate the most
    depleted zone whose window is open, whenever depletion exceeds the
    readily-available-water threshold $\rho\cdot\mathrm{TAW}$
    \cite{allen1998fao56}, subject to the budget and one offtake per day.
    This mimics the decision process of a local farm manager.
    \item \textbf{Simulated annealing (SA)} \cite{kirkpatrick1983sa}:
    single-flip Metropolis on the full QUBO with geometric cooling;
    the initial temperature is calibrated to the mean uphill move size of
    each instance.
    \item \textbf{Genetic algorithm (GA)} \cite{goldberg1989ga}: the
    standard metaheuristic of the classical irrigation-scheduling
    literature \cite{zhang2022canal,nguyen2017irrigation}; population 50,
    tournament selection (size 3), uniform crossover, bit-flip mutation at
    rate $1/n$, elitism of 2.
    \item \textbf{Exact MIQP}: branch-and-bound with presolve and cutting
    planes on the hard-constrained formulation
    $\min_x H_{\mathrm{obj}}(x)$ subject to $\sum_i x_i\le K$,
    $x\in\{0,1\}^{|\mathcal{W}|}$ (Gurobi~13.0 \cite{gurobi}, falling back
    to SCIP~10.0 \cite{scip} where the available license's size limit
    binds), whose optimum coincides with the penalized-QUBO optimum by
    Proposition~\ref{prop:penalty}; agreement with exhaustive enumeration
    is verified at $n=11$ and $n=21$. This solver both certifies ground
    truth for the scaling ladder and represents the practical classical
    frontier for deployment.
\end{enumerate}
SA and GA operate on the same $n$-bit search space (decision plus slack
bits) as the quantum solvers and receive an identical budget of
$2\times10^4$ objective evaluations, repeated over 20 seeds; the scaling
study additionally reports GA at a $10\times$ budget to separate
budget-division effects from genuine hardness (Section~\ref{sec:scaling}).

\subsection{Quantum Pipeline}
\label{sec:quantum-methods}
The QAOA ansatz \cite{farhi2014qaoa} alternates the diagonal cost unitary
$e^{-i\gamma_k H_C}$ (see \eqref{eq:ising}, coefficients normalized by
$\max(\max_i|h_i|,\max_{ij}|J_{ij}|)$) with the transverse-field mixer
$e^{-i\beta_k \sum_i X_i}$, applied to $\ket{+}^{\otimes n}$ at depths
$L\le 5$ (small instance) and $L\le 2$ (medium). Because $H_C$ is diagonal, ideal simulation is performed with an
exact statevector propagator that applies cost layers as elementwise
phases; we verified it reproduces the compiled Qiskit circuit
\cite{qiskit2024} amplitude-for-amplitude. Variational parameters are
optimized with COBYLA \cite{powell1994cobyla} from a linear-ramp
(annealing-inspired) initialization \cite{zhou2020qaoa} plus five random
restarts.

Noisy execution follows the \emph{parameter transfer} paradigm
\cite{galda2021transfer}: parameters optimized on the ideal simulator are
frozen, and only sampling is performed on noisy devices, avoiding
hardware-in-the-loop optimization overhead. Noise-model simulation uses
Qiskit Aer with the calibrated noise model (basis gates, coupling map,
$T_1/T_2$, and readout errors) of \texttt{FakeFez}, a 156-qubit IBM Heron
r2 processor snapshot of the same architecture as the
\texttt{ibm\_fez} device used for all hardware runs, so that noise-model
and hardware results are architecturally comparable; transpilation is at
optimization level 3, and a noiseless shot-based run of the \emph{same
transpiled circuit} isolates sampling noise from device noise. Hardware
runs execute on \texttt{ibm\_fez} via the Qiskit Runtime
\texttt{SamplerV2} with XY4 dynamical decoupling \cite{viola1999dd} and
gate/measurement twirling \cite{wallman2016twirling}, 4096 shots.

\subsection{Constraint-Preserving XY-Mixer Variant}
\label{sec:xy-methods}
Beyond the standard transverse-field ansatz, we derive and evaluate a
constraint-preserving variant in the quantum alternating operator ansatz
framework \cite{hadfield2019qaoa,wang2020xy}. Its full
derivation is given in Appendix~\ref{app:xy}. In the scarcity regime the
budget binds --- we verify by exhaustive enumeration that the small- and
medium-instance optima contain exactly $K$ irrigation events --- so the
search may be restricted to the Hamming-weight-$K$ manifold
$B_K=\{x\in\{0,1\}^{|\mathcal{W}|}:\sum_i x_i=K\}$. (This premise must be
checked per instance: the large-tier optimum uses only 6 of $K=8$ events,
so there the restriction would require a weight ladder; see
Appendix~\ref{app:xy}.) The cost Hamiltonian is then
$H_{\mathrm{obj}}$ alone: the budget penalty \emph{and its slack register
disappear}, shrinking 11 qubits to 9 (small) and 21 to 18 (medium) and
collapsing the interaction graph from near-complete to the physical
couplings only. The mixer is an ordered product of two-qubit XY rotations
$U_{ij}(\beta)=\exp[-i\beta(X_iX_j+Y_iY_j)/2]$ over a brickwork ring,
each of which exchanges amplitude only between equal-weight bitstrings,
so the evolution never leaves $B_K$; the initial state is the Dicke state
$\ket{D^{|\mathcal{W}|}_K}$ (uniform over $B_K$), prepared with the
deterministic construction of B\"artschi and Eidenbenz
\cite{bartschi2019dicke}. The ideal simulator applies exactly the same
ordered product as the compiled circuit, so no Trotter approximation is
involved at any depth; we verified simulator and circuit agree
amplitude-for-amplitude with zero leakage out of $B_K$. Hardware
execution uses the identical protocol as the standard ansatz (parameter
transfer, XY4 dynamical decoupling, gate and measurement twirling, 4096
shots). Because hardware noise \emph{does} break weight conservation,
samples that leak out of $B_K$ are scored with the penalized Hamiltonian
at the optimal slack setting,
$H_{\mathrm{obj}}+\lambda_B\max(0,\textstyle\sum_i x_i-K)^2$, the same
energy scale as the standard-ansatz metrics.

\subsection{Metrics}
Solution quality is reported as the normalized approximation ratio
$r=(E_{\max}-E)/(E_{\max}-E_{\min})\in[0,1]$ computed against the
enumerated spectrum extrema, together with the relative optimality gap
$(E-E_{\min})/|E_{\min}|$, the probability $P_{\mathrm{opt}}$ of sampling
the exact optimizer, the mean best-of-4096-shots energy, and the feasible
(budget-respecting) fraction of samples. For sampled distributions
$E=\langle H\rangle$ is the empirical mean energy.

%% file: sec_results.tex
\section{Results}
\label{sec:results}

\subsection{Energy Landscape and Certified Feasibility}
\begin{figure}[t]
    \centering
    \includegraphics[width=\linewidth]{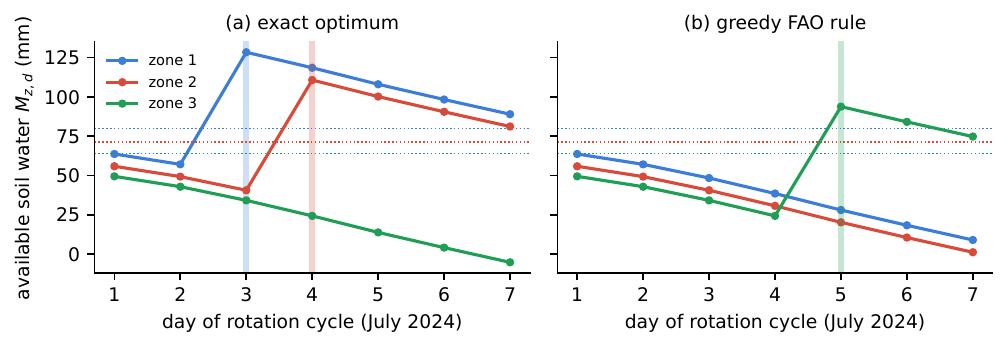}
    \caption{Root-zone available water trajectories for the small
    instance, driven by observed 2024 Khorezm forcing; dotted lines mark
    the targets $T_z$ and shaded verticals the irrigation events. (a)
    Exact optimum: triage --- zones 1 and 2 are dosed before their windows
    close and zone 3 is deliberately sacrificed. (b) Greedy FAO-trigger
    rule: the depletion trigger fires only after the windows of zones 1
    and 2 have closed; one dose goes to zone 3 and the second is never
    spent, leaving a $48.7\%$ optimality gap.}
    \label{fig:schedule}
\end{figure}

Exhaustive enumeration of the small ($n=11$) and medium ($n=21$) instances
confirms the picture painted in Section~\ref{sec:formulation}: the spectrum
spans a factor $\sim$59 (small) between the optimum
($E_{\min}=10492$) and the worst configuration, with the vast majority
of the hypercube lifted by the budget penalty; feasible schedules occupy a
narrow band at the bottom. In both instances the enumerated global optimum
satisfies the water budget exactly, validating the certified penalty
weight of Proposition~\ref{prop:penalty} at
$\lambda_B=5.4\times10^{3}$ (small) and $4.7\times10^{4}$ (medium) ---
$47$--$50\times$ smaller than the generic range prescription would
demand. The optimal small-tier schedule (Fig.~\ref{fig:schedule}a) performs
explicit triage: it irrigates zone 1 on the final day of its canal window
and zone 2 mid-window, deliberately sacrificing zone 3 --- the zone with
the smallest water-holding capacity --- to scarcity. Both doses are timed
as late as moisture memory rewards, anticipating window closure.

\subsection{Classical Baselines}
Table~\ref{tab:results} summarizes solver performance. The greedy FAO-rule
scheduler --- a proxy for current operational practice --- is
feasible but leaves optimality gaps of $48.7\%$ (small), $110.8\%$
(medium), and $98.2\%$ (large). Its failure mode on the small instance
(Fig.~\ref{fig:schedule}b) is instructive: the depletion trigger fires
only \emph{after} stress accumulates, by which time the canal windows of
zones 1 and 2 have already closed; the rule then spends one dose on zone 3
and strands the second dose entirely, starving two zones. Reactive
trigger scheduling is structurally blind to rotational window closure ---
precisely the coupling the QUBO encodes. Under a matched budget of $2\times10^4$
objective evaluations, both SA and GA recover the exact optimum in every
seed on the small instance. On the medium instance SA finds the optimum in
only $40\%$ of seeds (mean gap $2.1\%$) while GA remains at $100\%$; on
the large 40-variable instance GA attains a $0.46\%$ mean gap against the
proven optimum while SA stalls at $62.9\%$ --- confirming GA as a
strong fixed-budget heuristic for this problem family, consistent
with the prominence of evolutionary metaheuristics in the irrigation-scheduling literature
\cite{zhang2022canal,nguyen2017irrigation}. All heuristic runs complete in
tens of milliseconds, and the exact MIQP solver proves global optimality
of all three base tiers in at most $6.3$ s (Section~\ref{sec:scaling}).

\subsection{Classical Scaling Study}
\label{sec:scaling}
\begin{figure}[t]
    \centering
    \includegraphics[width=\linewidth]{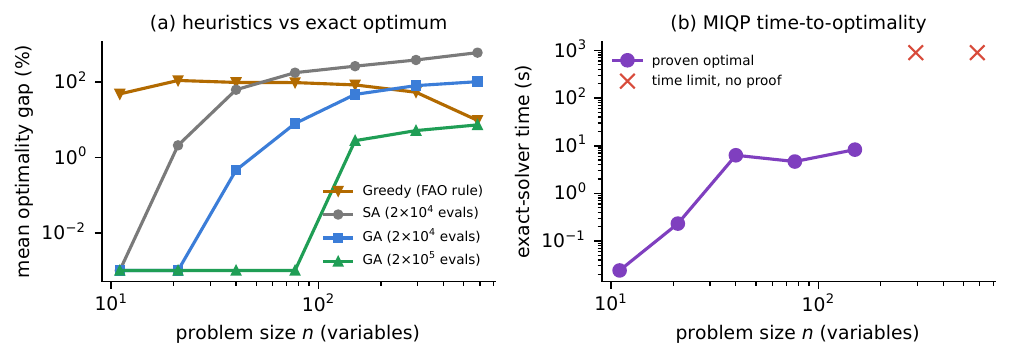}
    \caption{Classical scaling study on the real-soil instance ladder.
    (a) Mean optimality gap (20 seeds) of matched-budget SA and GA
    ($2\times10^4$ evaluations), GA at $10\times$ budget, and the greedy
    operational rule, relative to the proven optimum ($n\le150$) or the
    best-known solution ($n=295,584$). (b) Exact MIQP
    time-to-optimality; crosses mark the license-capped fallback solver
    hitting its 900-s limit without a proof. Fixed-budget heuristic
    degradation is a budget artifact (the $10\times$-budget GA restores
    near-optimality), and certified exact solution remains fast through
    $n=150$: no classical wall is observed at these sizes.}
    \label{fig:scaling}
\end{figure}

To test the classical-intractability narrative directly rather than by
citation, we extend the instance family along the ladder of
Section~\ref{sec:instances} ($n=40,77,150,295,584$ variables, all built from
real SoilGrids soil at 24 field sites and observed 2024 forcing) and run
the matched-budget heuristics against the exact MIQP frontier
(Fig.~\ref{fig:scaling}). Three facts emerge. First, branch-and-bound
proves \emph{global optimality in seconds} well past any width reachable
by near-term quantum hardware: $6.3$ s at $n=40$, $4.7$ s at $n=77$,
$8.3$ s at $n=150$ (Gurobi 13.0 \cite{gurobi}). Beyond $n=150$ the
binding constraint is our solver \emph{license}, not the problem: the
size-capped Gurobi build rejects larger quadratic models, and the
open-source fallback (SCIP 10.0 \cite{scip}) exhausts a 900-s budget at
$n=295$ and $584$ without a proof --- its incumbents are in fact
overtaken by the $10\times$-budget GA (by $18\%$ at $n=295$ and
$2.3\times$ at $n=584$) and its dual bounds remain uninformative for the
unpresolved nonconvex QUBO, so gaps at the two largest rungs are
reported against the best-known solution and we draw no hardness
conclusion from a tooling ceiling.
Second, the \emph{fixed}-budget heuristics do degrade with size --- SA
from a $0\%$ to a $265\%$ mean gap and GA from $0\%$ to $47\%$ between
$n=11$ and $n=150$ --- but this is an evaluation-budget artifact, not
hardness: at $10\times$ the budget ($1$--$73$ s per seed) GA returns to a
$2.8\%$ mean gap at $n=150$, recovers the proven optimum at every smaller
rung, and stays within $5.2\%$ ($n=295$) and $7.3\%$ ($n=584$) of
best-known at the uncertified rungs. Third, the deterministic greedy rule
leaves $84$--$111\%$ gaps across the measured-data tiers; its relative
gap narrows on the largest shared-forcing rungs ($54\%$ at $n=295$,
$10\%$ at $n=584$, against best-known references that are themselves
unproven there), reflecting the growing parameter homogeneity of those
synthetic-extension instances. The value of \emph{any} optimizer over
trigger-rule practice remains large at operational sizes, while the value
of exotic optimizers over commodity MIQP is nil at every certified size. We
therefore state plainly: \textbf{no classical scalability wall is
observed for this problem family at deployment-relevant sizes}, and the
contribution of this work is the physically grounded, hardware-ready
encoding with a certified constraint treatment and a transparent
benchmark --- not a demonstration of classical intractability.

\subsection{Ideal QAOA}
Before quoting optimum-sampling statistics we fix the baseline:
the small instance has only $2^{11}=2048$ configurations, so
\emph{uniform random sampling} finds its unique optimum within 4096 shots
with probability $1-(1-2^{-11})^{4096}\approx 0.86$. Sampling the $n=11$
optimum is therefore nearly guaranteed for any sampler and is \emph{not}
evidence of algorithmic capability; the small tier exists to validate the
end-to-end pipeline (encoding, transpilation, error mitigation,
readout), and the meaningful quantity at every width is the
\emph{enrichment factor} $P_{\mathrm{opt}}/2^{-n}$ --- how much
probability the ansatz concentrates on the optimizer relative to uniform
--- together with the feasible fraction and the best-of-shots gap. On the
medium instance ($2^{21}\approx 2.1\times10^{6}$ configurations), where
uniform 4096-shot sampling would find the optimum with probability only
$\approx 0.2\%$, these metrics are genuinely informative.

\begin{figure}[t]
    \centering
    \includegraphics[width=\linewidth]{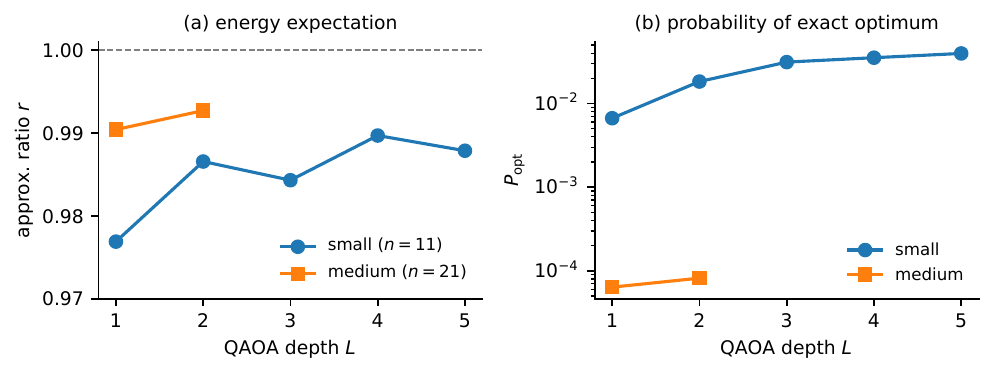}
    \caption{Ideal QAOA depth scan. (a) Normalized approximation ratio of
    the energy expectation; dotted/dashed gray lines mark the SA mean over
    20 matched-budget seeds for the small/medium instances. (b)
    Probability of sampling the exact optimum; uniform baselines are
    $2^{-11}$ and $2^{-21}$.}
    \label{fig:depth_scan}
\end{figure}

Figure~\ref{fig:depth_scan} shows the ideal (exact statevector)
performance. The energy expectation
$\langle H\rangle$ improves with depth ($r=0.977$ at $L=1$ to
$r\approx0.99$ at $L=4$--$5$ on the small instance) but its optimality
gap remains large
($59$--$133\%$) because the expectation averages over the residual
amplitude in the penalty band --- a generic feature of penalized QUBOs
executed at shallow depth. Operationally, however, QAOA is a
\emph{sampler}. On the
small instance $P_{\mathrm{opt}}$ grows monotonically from $0.67\%$
($L=1$) to $3.97\%$ ($L=5$), i.e.\ an enrichment of $14$--$81\times$ over
uniform, and the mean best-of-4096-shots energy sits at a numerically
zero gap at every depth --- consistent with, but not proving more than,
the $86\%$ uniform baseline above. On the medium instance the enrichment
is the substantive result: $P_{\mathrm{opt}}$ reaches
$6.4\times10^{-5}$ at $L{=}1$ and $8.2\times10^{-5}$ at $L{=}2$ ---
$134\times$ and $172\times$ the uniform baseline $2^{-21}$ --- and
best-of-4096 sampling lands within $2.2$--$2.3\%$ of the optimum
(Table~\ref{tab:results}), with all metrics improving monotonically in
depth.

\subsection{Noise-Model Execution}
\begin{figure}[t]
    \centering
    \includegraphics[width=\linewidth]{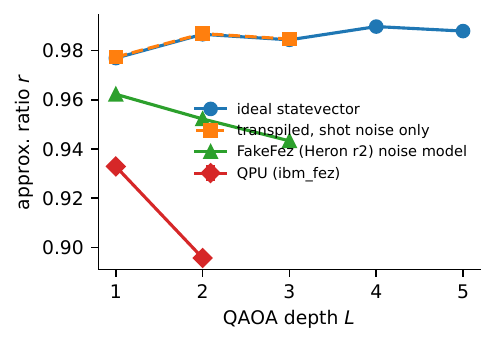}
    \caption{Approximation ratio of the sampled energy on the small
    instance: ideal statevector, transpiled circuits under shot noise
    only, the calibrated \texttt{FakeFez} (Heron r2) noise model, and IBM
    Heron hardware (\texttt{ibm\_fez}; the $L{=}1$ point is the mean of
    four repeat executions). Noise inverts the benefit of depth.}
    \label{fig:noise}
\end{figure}

Under the calibrated \texttt{FakeFez} noise model (a Heron r2 snapshot
architecturally matched to the \texttt{ibm\_fez} hardware used below;
Fig.~\ref{fig:noise}), transpiling the dense interaction graph onto
heavy-hex connectivity costs $224$, $449$, and $623$ two-qubit gates at
$L=1,2,3$, and accumulated gate error outweighs the algorithmic benefit
of depth in the energy expectation: the sampled ratio falls from
$r=0.962$ ($L=1$) to $r=0.943$ ($L=3$) while the noiseless-transpiled
reference of the same circuits holds at $r\approx0.98$. The feasible
fraction of samples decays from $59.2\%$ to $46.8\%$ ($6.6\times$ to
$5.2\times$ the uniform $9.0\%$), reflecting the uniformizing pull of
depolarizing noise toward the budget-violating bulk of the hypercube.
The optimum-sampling signal survives at all tested depths:
$P_{\mathrm{opt}}=0.51$--$0.76\%$, an enrichment of $11$--$16\times$ over
uniform. On the medium instance at $L{=}1$ ($711$ two-qubit gates) the
model predicts $r=0.971$, a $38.5\%$ feasible fraction, a best-of-4096
gap of $8.3\%$, and zero optimum counts in 4096 shots --- bracketing the
hardware observations below: the measured best-shot gap matches the
model, while measured feasibility is $\approx2\times$ lower, consistent
with drift and crosstalk absent from the static snapshot.

\subsection{Quantum Hardware}
\label{sec:hardware}
On \texttt{ibm\_fez} (IBM Heron, 156 qubits), with XY4 dynamical
decoupling, gate/measurement twirling, and parameters transferred from
the ideal optimization, the 11-qubit instance transpiles to $224$
($L{=}1$) and $449$ ($L{=}2$) two-qubit gates. Across four repeat
executions at $L{=}1$ spanning calibration drift,
$P_{\mathrm{opt}}=0.26\pm0.04\%$ --- an enrichment of
$5.2\pm0.8\times$ over uniform ($7.0\times$ in the single $L{=}2$ run,
statistically compatible) --- while $44.8\pm0.9\%$ ($L{=}1$) and
$31.0\%$ ($L{=}2$) of shots return budget-feasible schedules. Mean
energies sit $\approx2\times$ further from the optimum than the
noise-model prediction (penalty-band scatter dominates the expectation),
consistent with residual crosstalk and drift not captured by the static
snapshot. As argued above, sampling the $n{=}11$ optimum itself is close
to guaranteed at this state-space size and we do not report it as a
capability; the pipeline-level conclusion is the persistent, repeatable
enrichment factor.

The substantive hardware result is the 21-qubit instance, which we
execute on \texttt{ibm\_fez} at $L=1$ (three repeats; $711$ two-qubit
gates after routing) and $L=2$ ($1376$). Here the uniform baseline is
vanishing: $12288$ uniform shots would contain the optimum with
probability only $0.58\%$. The hardware sampled the exact optimum once
across the three $L{=}1$ runs ($P_{\mathrm{opt}}\approx 8\times10^{-5}$
pooled, a point estimate $\approx170\times$ uniform, though a single
count carries a wide Poisson interval), rejecting the uniform-sampling
null at $p\approx0.006$ and consistent with the ideal-parameter value
$6.4\times10^{-5}$ attenuated by circuit fidelity. Robust across all
repeats: $16.9\pm0.6\%$ of shots are budget-feasible ($10.9\times$ the
uniform feasible fraction of $1.5\%$), and the best-of-4096 shot lands
$8.4$--$9.0\%$ from the optimum at $L{=}1$ ($8.1\%$ at $L{=}2$; $0\%$ in
the repeat that sampled the optimizer). Depth again does not pay under
noise: $L{=}2$ halves the feasible fraction ($6.3\%$) without improving
the best shot. All job identifiers, calibration timestamps, and per-run
details are listed in Appendix~\ref{app:experiments}.

\subsection{The XY-Mixer Variant: Where It Helps and Where It Does Not}
\label{sec:xy}
The measured failure modes above --- penalty-band spectral compression
and SWAP overhead from the budget clique --- both trace to the artificial
budget term, and the constraint-preserving variant of
Section~\ref{sec:xy-methods} removes exactly that term. Evaluating it
end-to-end yields a deliberately two-sided verdict.

\emph{Ideal.} On the weight-$K$ manifold the small instance has only
$\binom{9}{2}=36$ feasible states, so its Dicke initial state is already
within $21\%$ of the optimum in expectation and uniform-in-subspace
sampling is itself a strong baseline; XY-QAOA concentrates little beyond
it ($P_{\mathrm{opt}}$ stays at $\approx 1\times$ the $1/36$ baseline for
$L\le3$) --- an honest null result at this width. The medium instance
($\binom{18}{4}=3060$ states) is more informative: $P_{\mathrm{opt}}$
reaches $0.31\%$ at $L{=}2$, a $9.6\times$ enrichment over the Dicke
baseline and a $38\times$ larger per-shot optimum probability than the
standard ansatz at equal depth, with every ideal sample feasible by
construction and the best-of-4096 gap numerically zero.

\emph{Resources.} At fixed width the variant is cheaper per layer ---
the cost graph collapses to the physical couplings --- but pays a fixed
Dicke-preparation overhead. On the small instance the trade is
favorable at every depth: $181/253/323$ two-qubit gates at $L=1/2/3$
versus $224/449/623$ for the standard ansatz, on $9$ versus $11$ qubits.
On the medium instance the preparation overhead ($\approx840$ two-qubit
gates at $n_x=18$, $K=4$) dominates: $1182$ versus $711$ gates at
$L{=}1$, reaching parity only near $L\approx2.5$, and the sequential
structure of preparation plus ring mixer multiplies the two-qubit
\emph{depth} by $3$--$5\times$ at equal $L$.

\emph{Hardware.} Both effects print through on \texttt{ibm\_fez}. On the
small instance the XY circuits sample the exact optimum at every depth
with $P_{\mathrm{opt}}=0.64$--$0.93\%$ --- $2.5$--$3.6\times$ the
standard ansatz's hardware rate at comparable or lower gate count, and a
mean-energy gap roughly half the standard ansatz's --- though still below
the $2.8\%$ Dicke baseline, and with the ideal feasibility guarantee
visibly broken by noise ($38\%$ of shots remain in $B_K$ at $L{=}1$,
decaying to $23\%$ at $L{=}3$; single $X$ errors leave the weight
manifold). On the medium instance the preparation overhead is fatal on
current hardware: $2.0\%$ ($L{=}1$) and $1.3\%$ ($L{=}2$) of shots remain
feasible, the optimum is never sampled, and the best feasible shot lands
$2.5\%$ ($L{=}1$) and $13.6\%$ ($L{=}2$) from the optimum. The variant
therefore delivers a real advantage exactly where its overhead is
amortized (small width, moderate depth) and is noise-dominated where the
Dicke preparation exceeds the budget-penalty routing it replaces --- a
quantitative design rule for constraint-preserving mixers on
heavy-hex devices, reported here in both directions. We also note the
structural caveat of Appendix~\ref{app:xy}: the weight restriction is
exact only when the budget binds, which holds for the small and medium
instances but not for the large tier.

\subsection{Synthesis}
At $11$--$21$ qubits no quantum advantage exists, and none is claimed:
the exact MIQP solver proves optimality in milliseconds-to-seconds at
every width we can embed on hardware, and through $7\times$ that width
(Section~\ref{sec:scaling}). The value of the demonstration is
structural. First, the full physical problem --- moisture memory, spatial
coupling, rotational windows, water budget --- executes end-to-end on
present-day hardware with a repeatable optimum-enrichment signal
($5\times$ uniform at $n{=}11$ over four runs; a nonzero optimum count
consistent with $\sim10^{2}\times$ uniform at $n{=}21$) and an
order-of-magnitude feasibility enrichment. Second, the measured failure
modes are informative: the dominant cost is not decoherence per se but
the interaction of (i) penalty-band spectral compression and (ii) SWAP
overhead from embedding the budget clique in heavy-hex connectivity.
Third, the paper's own proposed remedy --- the constraint-preserving XY
mixer --- was built and measured: it
verifiably removes both failure modes and improves every per-gate metric
at small width, while its state-preparation overhead defines the width
frontier beyond which it stops paying on current devices.

\input{tab_results}

%% file: tab_results.tex
\begin{table}[t]
\centering
\caption{Solver comparison on the observed data Khorezm instances. Gap is the
relative optimality gap $(E-E_{\min})/|E_{\min}|$ in percent: ``mean'' uses
the mean energy over 20 seeds (classical) or the energy expectation
$\langle H\rangle$ (QAOA); ``best'' uses the best seed (classical), the
mean best-of-4096-shots energy (ideal QAOA), or the best sampled shot
(noisy/QPU). $P_{\mathrm{opt}}$ is the per-seed probability of hitting the
exact optimum (classical) or the per-shot probability of sampling it
(QAOA). Classical heuristics use $2\times10^4$ objective evaluations and
run in tens of milliseconds; QAOA rows use the depth with the best
$\langle H\rangle$ (XY QPU: best $P_{\mathrm{opt}}$). QPU rows marked
$\times k$ average $k$ repeat executions spanning calibration drift; XY
rows use the constraint-preserving mixer of Section~\ref{sec:xy-methods},
with leaked hardware samples scored at optimal slack
(Appendix~\ref{app:xy}).}
\label{tab:results}
\begin{tabular}{lccc}
\toprule
Solver & Gap$_{\mathrm{mean}}$ (\%) & Gap$_{\mathrm{best}}$ (\%) & $P_{\mathrm{opt}}$ \\
\midrule
\multicolumn{4}{l}{\emph{small} instance, $n=11$, $E_{\min}=10492$}\\[1pt]
Greedy (FAO rule) & 48.69 & --- & ---\\
SA & 0.00 & 0.00 & 1.00\\
GA & 0.00 & 0.00 & 1.00\\
QAOA ideal ($L{=}4$) & 59.41 & 0.00 & $0.035$\\
QAOA noisy ($L{=}1$) & 217.78 & 0.00 & $0.005$\\
QAOA QPU ($L{=}1$) $\times$4 & 386.70 & 0.00 & $0.003$\\
QAOA QPU ($L{=}2$) & 600.89 & 0.00 & $0.003$\\
XY ideal ($L{=}3$) & 13.25 & 0.00 & $0.025$\\
XY QPU ($L{=}1$) & 206.88 & 0.00 & $0.009$\\
\addlinespace
\multicolumn{4}{l}{\emph{medium} instance, $n=21$, $E_{\min}=15250$}\\[1pt]
Greedy (FAO rule) & 110.77 & --- & ---\\
SA & 2.10 & 0.00 & 0.40\\
GA & 0.00 & 0.00 & 1.00\\
QAOA ideal ($L{=}2$) & 773.50 & 2.19 & $8.2\!\times\!10^{-5}$\\
QAOA noisy ($L{=}1$) & 3085.52 & 8.35 & $0$\\
QAOA QPU ($L{=}1$) $\times$3 & 8648.70 & 5.78 & $8.1\!\times\!10^{-5}$\\
QAOA QPU ($L{=}2$) & 13622.76 & 8.11 & $0$\\
XY ideal ($L{=}2$) & 174.70 & 0.00 & $0.003$\\
XY QPU ($L{=}1$) & 8803.69 & 2.53 & $0$\\
\addlinespace
\bottomrule
\end{tabular}
\end{table}

%% file: sec_limitations.tex
\section{Discussion and Limitations}
\label{sec:limitations}

\textbf{Hydrological model error.} Assumption~\ref{as:linear} discards
saturation clipping: a schedule that overfills a zone would, in reality,
lose water to deep percolation rather than accumulate a quadratic stress
penalty. The two mechanisms agree in sign (over-irrigation is bad) but not
in magnitude, so the QUBO objective is a surrogate rather than an exact
water-balance simulator. In the scarcity-dominated regime studied here
(budget $K$ well below the unconstrained irrigation demand) optimal
schedules stay below field capacity, so the \emph{upper} clipping bound
never binds. The lower bound is a different matter: the optimal
small-instance schedule deliberately sacrifices zone~3, driving it toward
and below the wilting point (Fig.~\ref{fig:schedule}a) --- the
sub-zero-deficit regime where the unclipped linear bucket is \emph{least}
faithful, since real infiltration, root uptake, and capillary rise all
change character near wilting. The approximation is therefore benign for
the zones the optimizer protects, but only qualitative for the zone it
triages away; validating optimized schedules through a full
Richards-equation or HYDRUS-1D simulation, as calibrated for Khorezm by
Forkutsa et al. \cite{forkutsa2009modeling}, remains future work.

\textbf{Symmetric stress surrogate.} The quadratic
$(M-T)^2$ stress term penalizes surplus and deficit moisture equally,
whereas crop water-production functions are strongly asymmetric: FAO-56
practice applies a piecewise stress coefficient $K_s$ that activates only
once depletion exceeds the readily-available-water threshold, and
over-wetting damages yield through different (aeration, leaching)
mechanisms at different rates \cite{allen1998fao56,doorenbos1979yield}.
The symmetric square is the price of remaining 2-local: a piecewise or
one-sided penalty in the affine moisture state would introduce
higher-order or non-polynomial terms requiring auxiliary qubits. Together
with linearization (Assumption~\ref{as:linear}) these are the two modeling
concessions made for quantum compatibility; neither is specific to QAOA
--- any QUBO/Ising sampler inherits both.

\textbf{Data granularity.} No public archive provides daily offtake logs
for Uzbek WCAs, so canal windows, doses, and the initial depletion state
are \emph{scenario} parameters grounded in documented practice
\cite{adb2020uzbekistan,veldwisch2008cotton} rather than measured
schedules; the meteorological forcing (NASA POWER) has a
$0.5^\circ\times0.625^\circ$ native grid that assigns the same weather to
all three zones, and SoilGrids carries its own mapping uncertainty
\cite{soilgrids}. None of these caveats affect the structure of the QUBO
since they shift coefficients, not couplings.

\textbf{Penalty landscape compression.} Hard constraints enter through
$\lambda_B$, and although Proposition~\ref{prop:penalty} keeps
$\lambda_B$ an order of magnitude tighter than the generic prescription,
the budget-violating majority of the hypercube still dominates the
spectrum: near-optimal feasible states are compressed into a narrow energy
band. This compresses the standard approximation ratio toward unity
(motivating the optimality-gap metric of Section~\ref{sec:methods}) and, on
hardware, means uniform bit-flip noise preferentially scatters samples
into the high-penalty region.

\textbf{NISQ depth and width.} The budget term couples every decision
variable to the slack register, producing a near-complete interaction
graph whose $R_{ZZ}$ layers route poorly on heavy-hex topologies; the
transpiled two-qubit depth grows quickly with $L$, bounding useful depths
to $L\lesssim 2$ on current devices for our 11-qubit instance. The
constraint-preserving XY variant removes this term, but its
Dicke-preparation overhead defines its own width frontier
(Section~\ref{sec:xy}): neither ansatz escapes the depth budget of current
devices beyond $n\approx 20$. Statevector
optimization above $n\approx 25$ is likewise impractical classically, so
variational parameters for larger instances would require transfer or
extrapolation schemes \cite{galda2021transfer} rather than direct
optimization.

\textbf{Classical frontier, measured.} The scaling study of
Section~\ref{sec:scaling} ran the practical classical frontier rather: branch-and-bound MIQP proves global optimality in seconds, and fixed-budget metaheuristic
degradation is repaired by scaling the budget. Any future claim of
quantum utility for this problem family must therefore be made against
commodity exact solvers under realistic time budgets --- including warm
starts and re-planning loops --- not against fixed-budget heuristics.
The scaling ladder itself has one synthetic element: zones beyond the
three measured fields reuse the zone-1 weather (one MERRA-2 cell) with per-site SoilGrids soil, so rung-to-rung heterogeneity is
soil-driven by construction.

%% file: sec_conclusion.tex
\section{Conclusion}
\label{sec:conclusion}

We have shown that rotational irrigation scheduling --- the daily
operational problem of Central Asian water management --- maps onto a
2-local Ising Hamiltonian once the soil-water balance is linearized: the
dominant structure is intra-zone temporal moisture memory, joined by
sparse same-day spatial couplings, while rigid canal rotations become
qubit-count reductions and only the water budget
requires an artificial penalty coupling. The formulation is built
end-to-end from observed data for a cotton district in Khorezm,
Uzbekistan (NASA POWER meteorology, SoilGrids soil hydraulics, FAO-56
agronomy, and documented WCA rotation practice), and comes with a
certified, instance-adaptive penalty weight that keeps the constraint
landscape an order of magnitude tighter than generic prescriptions.

The benchmark verdict is deliberately symmetrical and fully measured. On
the classical side, an off-the-shelf MIQP solver proves global optimality
in seconds on soil instances up to $7\times$ the width of our
largest hardware run (and license-capped open-source solving still
returns the best-known solutions at $28\times$), while fixed-budget
metaheuristic degradation vanishes when the budget scales: no classical
wall is observed at deployment-relevant sizes, and we claim none. On the quantum side, the full physical problem
executes end-to-end on IBM Heron hardware with repeatable
optimum-enrichment over uniform sampling ($5\times$ at 11 qubits across
four runs; a nonzero optimum count consistent with
$\sim10^{2}\times$ uniform at 21 qubits, where uniform sampling almost
surely would find nothing) and an order-of-magnitude feasibility
enrichment --- while the greedy rule that mimics local operational practice
leaves half to all of the attainable objective on the table on the
measured-data tiers. The instrumented failure modes (penalty-band compression, budget-
clique routing) motivated a constraint-preserving XY-mixer variant that eliminates
the budget penalty and its slack qubits, improves per-shot optimum
sampling and gate counts on hardware at small width, and is noise-dominated at 18 qubits,
where its Dicke-preparation overhead exceeds the routing cost it removes.

The contribution, then, is a hardware-ready, physically derived, and fully
reproducible encoding --- with certified constraint handling, an acknowledged
classical frontier, and a measured mixer-design rule --- positioned for
the regimes where the classical picture may genuinely change: tight re-planning loops, and
basin-scale networks. Closed-loop deployment, in which each rotation
cycle's QUBO is re-instantiated from fresh sensor and forecast data,
remains the goal that the Aral Sea basin's agriculture urgently needs;
the encoding presented here is ready for both classical and NISQ-era quantum hardwares.

%% file: app_derivation.tex
\appendices

\section{Complete QUBO Derivation}
\label{app:derivation}

\subsection{Expansion of the stress term}
Fix a zone $z$ and abbreviate $\chi_d:=\sum_{d'\le d,\,(z,d')\in\mathcal{W}}
x_{z,d'}$, so that $M_{z,d}-T_{z,d}=a\chi_d+\delta_{z,d}$ by
\eqref{eq:closedform}. Then
\begin{equation}
    \sum_{d=1}^{D} w_{z,d}\big(a\chi_d+\delta_{z,d}\big)^2
    = \sum_{d=1}^{D} w_{z,d}\Big[a^2\chi_d^2
    + 2a\,\delta_{z,d}\,\chi_d + \delta_{z,d}^2\Big].
\end{equation}
Using idempotency $x^2=x$ of binary variables,
\begin{equation}
    \chi_d^2 = \sum_{d'\le d} x_{z,d'}
    \;+\; 2\!\!\sum_{d'<d''\le d}\!\! x_{z,d'}x_{z,d''},
\end{equation}
where all sums run over indices in $\mathcal{W}$. Exchanging the order of
summation between the observation day $d$ and the decision days $d',d''$
(a decision on day $d'$ affects every observation day $d\ge d'$; a pair
$(d',d'')$ with $d'<d''$ affects every $d\ge d''$) yields
\begin{align}
    \sum_d w_{z,d}\,a^2\chi_d^2
    ={}& a^2\sum_{d'} S_{z,d'}\,x_{z,d'} \nonumber\\
    &+ 2a^2\!\!\sum_{d'<d''}\!\! S_{z,d''}\,x_{z,d'}x_{z,d''},\\
    \sum_d w_{z,d}\,2a\,\delta_{z,d}\chi_d
    ={}& 2a \sum_{d'} V_{z,d'}\,x_{z,d'},
\end{align}
with $S_{z,d'}=\sum_{d\ge d'} w_{z,d}$ and
$V_{z,d'}=\sum_{d\ge d'} w_{z,d}\,\delta_{z,d}$ as in the main text, giving
\eqref{eq:stress-qubo} with additive constant
$\sum_{z,d} w_{z,d}\,\delta_{z,d}^2$.

\subsection{QUBO coefficient tables}
Index the $n_x=|\mathcal{W}|$ decision variables by $i=(z,d)$ and the $m$
slack bits by $k$. Writing
$H=\sum_i \ell_i x_i + \sum_{i<j} q_{ij}x_i x_j + \mathrm{const}$, the
nonzero coefficients of \eqref{eq:total} are:
\begin{align}
    \ell_{(z,d')} &= c_w a + a^2 S_{z,d'} + 2a V_{z,d'}
        + \lambda_B(1-2K), \\
    \ell_{y_k} &= \lambda_B\, c_k(c_k - 2K), \\
    q_{(z,d'),(z,d'')} &= 2a^2 S_{z,\max(d',d'')} + 2\lambda_B
        + \lambda_T\,[\,|d'-d''|=1\,], \\
    q_{(z,d),(z',d)} &= \lambda_S\,[\,(z,z')\in\mathcal{E}\,] + 2\lambda_B, \\
    q_{(z,d),(z',d'')} &= 2\lambda_B \qquad (z\ne z',\, d\ne d''), \\
    q_{(z,d),y_k} &= 2\lambda_B\, c_k, \qquad
    q_{y_k,y_{k'}} = 2\lambda_B\, c_k c_{k'},
\end{align}
where $[\cdot]$ is the Iverson bracket and the slack coefficients are
$c_k=2^k$ for $k<m-1$, $c_{m-1}=K-2^{m-1}+1$ with
$m=\lceil\log_2(K{+}1)\rceil$, so that
$s(y)=\sum_k c_k y_k$ ranges exactly over $\{0,\dots,K\}$
\cite{lucas2014ising}. The budget expansion also contributes the constant
$\lambda_B K^2$.

\subsection{Non-uniform doses}
If zone doses $a_z$ differ, the budget constraint
$\sum_{z,d} a_z x_{z,d}\le B$ is first rescaled by the greatest common
divisor $g$ of $\{a_z\}$ (doses being set by hydromodule norms in multiples
of $10\,\mathrm{mm}$ in practice), giving integer coefficients
$u_z=a_z/g$ and budget $K=\lfloor B/g\rfloor$; the slack register then
requires $m=\lceil\log_2(K{+}1)\rceil$ qubits and the stress-term algebra
carries through with $a\to a_z$ throughout. All expressions remain 2-local.

\subsection{Proof of Proposition~\ref{prop:penalty}}
Suppose $(x,y)$ violates the budget, $N:=\sum_i x_i > K$. Since the slack
satisfies $s(y)\ge 0$, the penalty argument is $t:=N+s(y)-K\ge 1$. Remove
any active event $i$ (set $x_i\to 0$). The objective changes by
$-\ell_i-\sum_{j\ne i} q_{ij}x_j \le -\ell_i \le \max(0,-\min_i \ell_i)$,
using $q_{ij}\ge 0$; the penalty changes by
$\lambda_B\big[(t-1)^2-t^2\big]=-\lambda_B(2t-1)\le-\lambda_B$. Hence each
removal strictly lowers $H$ whenever
$\lambda_B>\max(0,-\min_i\ell_i)$. Iterating until $\sum_i x_i = K$ and
re-choosing the slack so that $s(y)=K-\sum_i x_i$ (possible since $s$
ranges over $\{0,\dots,K\}$, and $y$ does not appear in
$H_{\mathrm{obj}}$) produces a strictly better feasible configuration.
Therefore no budget-violating configuration is a global minimizer.
\hfill$\square$

The same argument applied to a feasible $x$ with misconfigured slack
($t\neq 0$) shows the optimal $y$ always zeroes the penalty. We set
$\lambda_B=1.1\times\max(1,-\min_i\ell_i)$ per instance.

\subsection{Ising mapping}
Substituting $x_i=(1-Z_i)/2$ into
$H=\sum_i \ell_i x_i+\sum_{i<j}q_{ij}x_ix_j+c$:
\begin{align}
    J_{ij} &= \frac{q_{ij}}{4}, \qquad
    h_i = -\frac{\ell_i}{2} - \frac{1}{4}\sum_{j\ne i} q_{ij}, \\
    E_0 &= c + \frac{1}{2}\sum_i \ell_i + \frac{1}{4}\sum_{i<j} q_{ij}.
\end{align}
Every term of \eqref{eq:total} is at most quadratic in $x$, hence
$H_C$ in \eqref{eq:ising} is exactly 2-local: the QAOA phase separator
$e^{-i\gamma H_C}$ compiles to $\binom{n}{2}$-bounded $R_{ZZ}$ rotations
(one per nonzero $J_{ij}$) and $n$ single-qubit $R_Z$ rotations, with no
Trotter error at any depth $L$. Before compilation we normalize $H_C$ by
$\max(\max_i|h_i|,\max_{i<j}|J_{ij}|)$, which rescales the variational
landscape without changing the minimizer.

\section{The Constraint-Preserving XY-Mixer Variant}
\label{app:xy}

This appendix derives the XY-mixer variant of Section~\ref{sec:xy-methods}.

\subsection{Restriction to the weight-$K$ manifold}
Let $B_K=\{x\in\{0,1\}^{n_x}:\sum_i x_i=K\}$ with
$\dim B_K=\binom{n_x}{K}$. The restriction of the search to $B_K$ is
exact if and only if the constrained optimum saturates the budget,
$\min_{x\in B_K}H_{\mathrm{obj}}(x)=\min_{\sum x\le K}H_{\mathrm{obj}}(x)$.
This premise is \emph{instance-dependent} and must be checked: we verify
it by exhaustive enumeration for the small ($\binom{9}{2}=36$ states) and
medium ($\binom{18}{4}=3060$ states) instances, whose penalized-QUBO
optima contain exactly $K$ events. It can fail when water is less scarce:
the large-tier optimum uses only $6$ of $K=8$ permitted events, so a
weight-restricted search there would require running the ansatz on a
ladder of manifolds $B_0,\dots,B_K$ (or an inequality-preserving mixer),
at $K{+}1$ times the sampling cost.

\subsection{The mixer preserves Hamming weight}
For an edge $(i,j)$ define
$U_{ij}(\beta)=\exp[-i\beta(X_iX_j+Y_iY_j)/2]$. In the two-qubit
computational basis, $(X_iX_j+Y_iY_j)/2$ annihilates
$\ket{00}$ and $\ket{11}$ and acts as the Pauli-$X$ operator on the
$\{\ket{01},\ket{10}\}$ block, so
\begin{equation}
    U_{ij}(\beta)\big|_{\{\ket{01},\ket{10}\}} =
    \begin{pmatrix} \cos\beta & -i\sin\beta\\
                    -i\sin\beta & \cos\beta \end{pmatrix},
\end{equation}
and $U_{ij}(\beta)$ is the identity on $\ket{00}$ and $\ket{11}$.
Equivalently, $[X_iX_j+Y_iY_j,\; \sum_k Z_k]=0$, so each factor --- and
therefore any ordered product of factors --- commutes with total Hamming
weight. The mixer layer used here is the brickwork ring product
\begin{equation}
    U_M(\beta)=\!\!\prod_{(i,j)\in\mathcal{R}}\!\! U_{ij}(\beta),\qquad
    \mathcal{R}=\text{even pairs, odd pairs, seam},
    \label{eq:xy-ring}
\end{equation}
over the cycle $0\to1\to\dots\to n_x{-}1\to0$. Since adjacent
transpositions generate the full symmetric group, repeated layers of
\eqref{eq:xy-ring} connect every pair of states in $B_K$: the ansatz
is ergodic on the feasible manifold \cite{hadfield2019qaoa,wang2020xy}.
The phase layer applies the diagonal $H_{\mathrm{obj}}$ (QUBO
coefficients of Appendix~\ref{app:derivation} with
$\lambda_B=0$ and no slack variables), normalized as in the standard
ansatz.

\subsection{Initial state and circuit identity}
The initial state is the Dicke state
$\ket{D^{n_x}_K}=\binom{n_x}{K}^{-1/2}\sum_{x\in B_K}\ket{x}$, prepared
with the deterministic, ancilla-free construction of B\"artschi and
Eidenbenz \cite{bartschi2019dicke} ($O(K n_x)$ two-qubit gates, depth
$O(n_x)$). Because the cost layer is diagonal and the mixer layer is
implemented \emph{as the ordered product} \eqref{eq:xy-ring} both in
the ideal simulator and in the compiled circuit
(\texttt{XXPlusYY} gates), simulator and hardware realize the same
unitary with no Trotter error at any depth. We verified numerically that
(i) the prepared state matches the uniform weight-$K$ superposition,
(ii) the full circuit statevector matches the subspace simulator with
overlap $>1-10^{-9}$, and (iii) amplitude leakage out of $B_K$ is
$<10^{-9}$.

\subsection{Scoring convention under hardware noise}
Physical noise (bit flips, amplitude damping, readout error) does not
respect weight conservation, so hardware histograms contain strings
outside $B_K$. For all reported XY metrics such samples are scored with
the penalized Hamiltonian at the optimal slack setting,
\begin{equation}
    E(x) = H_{\mathrm{obj}}(x)
    + \lambda_B\,\max\Big(0,\sum_i x_i - K\Big)^{2},
\end{equation}
which coincides with $\min_y H(x,y)$ over the slack register of the
standard formulation (the slack absorbs $\sum_i x_i<K$ exactly, cf.\ the
proof of Proposition~\ref{prop:penalty}). XY and standard-ansatz energies
are therefore reported on the same scale, and a leaked string can never
be scored better than a feasible one of equal objective value.

\section{Instance Construction from Field Data}
\label{app:instance}

Deterministic inputs entering $\Phi_{z,d}$ and $\delta_{z,d}$ are computed
from the data sources of Section~\ref{sec:methods} as follows.
Reference evapotranspiration $\mathrm{ET}^0$ is computed daily with the
FAO-56 Penman--Monteith equation \cite{allen1998fao56} from NASA POWER
(MERRA-2) 2-m temperature extrema, relative humidity, wind speed, and
all-sky surface shortwave irradiance \cite{nasapower}; crop
evapotranspiration is $\mathrm{ET}^c=K_c\,\mathrm{ET}^0$ with the FAO-56
single-coefficient cotton curve ($K_{c,\mathrm{ini}}=0.35$,
$K_{c,\mathrm{mid}}=1.18$, $K_{c,\mathrm{end}}=0.60$; stage lengths
30/50/60/55 days from mid-April sowing). Effective precipitation uses the
CROPWAT fixed-percentage method, $P^{\mathrm{eff}}=0.8P$
\cite{fao_cropwat}. Capillary rise is fixed at the mid-season mean
$G=1.5$ mm\,day$^{-1}$ measured for Khorezm cotton fields with water
tables at 0.9--1.6 m by Forkutsa et al. \cite{forkutsa2009modeling}.
Total available water per zone,
$\mathrm{TAW}_z=(\theta^{\mathrm{FC}}_z-\theta^{\mathrm{WP}}_z)Z_r$ with
$Z_r=1.0$ m, uses SoilGrids~2.0 volumetric water contents at $-33$ kPa and
$-1500$ kPa, thickness-weighted over 0--100 cm \cite{soilgrids}. Stress
weights $w_{z,d}$ equal the FAO-33 yield-response factor of the
phenological stage on day $d$ \cite{doorenbos1979yield}. The depletion
fraction is $\rho=0.65$ (FAO-56, cotton). Initial state
$M_{z,0}=\mathrm{TAW}_z-0.6\,\rho\,\mathrm{TAW}_z$ (60\% of readily available
water depleted) defines the scheduling scenario. Canal windows follow the
staggered offtake rotation documented for WCA-managed distributaries in
Khorezm \cite{adb2020uzbekistan,veldwisch2008cotton}, with adjacent windows
overlapping by one day due to canal capacity transitions.

%% file: app_experiments.tex
\section{Experimental Details}
\label{app:experiments}

\subsection{Instance parameters}
Table~\ref{tab:params} lists the parameters of the benchmark instances,
all derived from the data sources of Appendix~\ref{app:instance}. The
schedule-free deficits $\delta_{z,d}$ reach $-70$ mm by the end of the
small-instance horizon --- roughly one dose $a=80$ mm short per zone ---
so the budget $K=2$ forces genuine triage among the three zones, making the instance combinatorially meaningful.

\begin{table}[!t]
\centering
\caption{Benchmark instance parameters. Soil constants are
thickness-weighted SoilGrids~2.0 values at the three field locations;
windows are days within the horizon when each zone's offtake is open.}
\label{tab:params}
\begin{tabular}{lccc}
\toprule
 & small & medium & large \\
\midrule
horizon $D$ (days) & 7 & 14 & 28 \\
budget $K$ (events) & 2 & 4 & 8 \\
decision variables $|\mathcal{W}|$ & 9 & 18 & 36 \\
slack bits $m$ & 2 & 3 & 4 \\
total $n$ & 11 & 21 & 40 \\
$\lambda_S=\lambda_T$ & 2377 & 2789 & 2994 \\
$\lambda_B$ (Proposition~\ref{prop:penalty}) & $5.40\times10^3$ & $4.66\times10^4$ & $1.91\times10^5$ \\
\bottomrule
\end{tabular}
\end{table}

Common to all tiers: dose $a=80$ mm, water price $c_w=1.0$ (stress
units)/mm, depletion fraction $\rho=0.65$, root depth $Z_r=1.0$ m, capillary
rise $G=1.5$ mm\,day$^{-1}$, stress weights $w\in\{0.2,0.5\}$ (FAO-33
vegetative/flowering), $\mathrm{TAW}=\{118.4,105.6,95.1\}$ mm,
$M_{z,0}=\{72.2,64.4,58.0\}$ mm, $T_z=\{79.9,71.3,64.2\}$ mm for zones
$1$--$3$. Small-tier windows: $\mathcal{W}_1=\{1,2,3\}$,
$\mathcal{W}_2=\{3,4,5\}$, $\mathcal{W}_3=\{5,6,7\}$; medium/large repeat
the stagger each subsequent week. Zone adjacency is the path $1$--$2$--$3$
(fields along one distributary).

\subsection{Optimizer settings}
Ideal-simulation parameters are optimized per depth with COBYLA
(\texttt{rhobeg}=0.3, 200 iterations) from a linear-ramp initialization
$\gamma_k$ increasing and $\beta_k$ decreasing across layers
\cite{zhou2020qaoa}, plus five uniform-random restarts; the best of the
six runs is kept. Classical SA uses single-flip Metropolis with geometric
cooling over $2\times10^4$ steps, initial temperature calibrated to
$2\times$ the mean uphill move magnitude, final temperature $10^{-3}$ of
initial. GA uses population 50, tournament size 3, uniform crossover,
mutation rate $1/n$, elitism 2, for $2\times10^4$ evaluations. All
stochastic solvers report statistics over 20 seeds.

\subsection{Classical scaling ladder}
The scaling rungs of Section~\ref{sec:scaling} extend the base tiers as
$(Z,D,K)=(6,28,16)$, $(12,28,32)$, $(24,28,64)$, and $(24,56,128)$,
giving $n=77$, $150$, $295$, and $584$ total variables. Zones are grouped
into parallel three-zone branches with the base staggered windows; soil
constants use SoilGrids~2.0 at 24 field locations along the
Urgench--Khiva corridor (coordinates in
\texttt{data/soil\_scaling.json} of the repository), and all zones share
the observed zone-1 2024 forcing (single MERRA-2 cell). The exact MIQP
tier solves $\min H_{\mathrm{obj}}$ s.t.\ $\sum x\le K$ with Gurobi
13.0.2 \cite{gurobi} (SCIP 10.0 \cite{scip} at $n\ge295$, where the
available Gurobi license's size limit binds), 900-s limit per rung,
validated against exhaustive enumeration at $n=11$ and $21$. SA/GA use
seeds $0$--$19$ at $2\times10^4$ evaluations (GA additionally at
$2\times10^5$).

\subsection{Noise-model and hardware execution}
Noise simulation uses Qiskit Aer with the calibrated noise model of
\texttt{FakeFez} (156-qubit Heron r2 snapshot: basis-gate errors,
$T_1/T_2$ relaxation, and readout error), matching the architecture of
the \texttt{ibm\_fez} hardware backend, after transpilation at
optimization level 3 (fixed seed). Hardware execution used the 156-qubit
Heron processor \texttt{ibm\_fez} through Qiskit Runtime \texttt{SamplerV2}
on 10 July 2026, 4096 shots per circuit, with XY4 dynamical decoupling and
gate + measurement twirling enabled; variational parameters were
transferred from the ideal optimization without on-device tuning.
Table~\ref{tab:jobs} lists every QPU job with its transpiled resource
counts. Repeat executions of identical circuits sample calibration drift;
the two batches ran against calibrations of 19:18 and 19:28 local time
(UTC+9). Total billed quantum time for all revision jobs was
$\approx40$ s.

\begin{table*}[!t]
\centering
\caption{IBM Quantum jobs on \texttt{ibm\_fez} (all 4096 shots, XY4
dynamical decoupling, gate + measurement twirling, transpiled at
optimization level 3 with fixed seed). ``std'' is the transverse-field
ansatz of Section~\ref{sec:quantum-methods} on $n$ qubits including slack;
``XY'' is the constraint-preserving variant of Section~\ref{sec:xy-methods}
on the $n_x$ decision qubits only.}
\label{tab:jobs}
\begin{tabular}{llccc}
\toprule
ansatz/tier & job ID & $L$ & depth & 2q gates \\
\midrule
std small   & \texttt{d9831rgtcv6s73dljo10} & 1 & 366 & 224 \\
std small   & \texttt{d98coal2su3c739j4dlg} & 1 & 366 & 224 \\
std small   & \texttt{d98cohqf47jc73a7opp0} & 1 & 366 & 224 \\
std small   & \texttt{d98cop4qp3as739spnpg} & 1 & 366 & 224 \\
std small   & \texttt{d9831rotcv6s73dljo2g} & 2 & 655 & 449 \\
std medium  & \texttt{d98ckv2f47jc73a7ol1g} & 1 & 692 & 711 \\
std medium  & \texttt{d98cp1cqp3as739spo20} & 1 & 692 & 711 \\
std medium  & \texttt{d98cpad2su3c739j4f80} & 1 & 692 & 711 \\
std medium  & \texttt{d98ckvgtcv6s73dlvalg} & 2 & 1023 & 1376 \\
XY small    & \texttt{d98cpicqp3as739spoog} & 1 & 511 & 181 \\
XY small    & \texttt{d98clbif47jc73a7olgg} & 2 & 677 & 253 \\
XY small    & \texttt{d98cpj0tcv6s73dlvghg} & 3 & 842 & 323 \\
XY medium   & \texttt{d98cpjl2su3c739j4fk0} & 1 & 2473 & 1182 \\
XY medium   & \texttt{d98clc8tcv6s73dlvb4g} & 2 & 3029 & 1527 \\
\bottomrule
\end{tabular}
\end{table*}

\subsection{Software}
Python 3.14; Qiskit 2.5 with Qiskit Aer and \texttt{qiskit-ibm-runtime}
\cite{qiskit2024}; NumPy/SciPy/pandas for the classical stack; Gurobi
13.0.2 \cite{gurobi} and PySCIPOpt/SCIP 10.0 \cite{scip} for the exact
MIQP tier. The ideal
QAOA statevector propagator exploits the diagonality of $H_C$ (elementwise
cost phases + single-qubit mixer contractions) and was verified
amplitude-for-amplitude against the compiled Qiskit circuits; the XY
subspace simulator was likewise verified against the compiled Dicke-plus-
ring circuits (Appendix~\ref{app:xy}). All stochastic components log
their seeds; every figure and table regenerates from
\texttt{results/*.json} via the repository scripts.

%% file: sec_acknowledgments.tex
\section*{Acknowledgment}
The authors acknowledge the use of IBM Quantum services for this work.
Meteorological data were obtained from the NASA Langley Research Center POWER Project,
and soil data from the ISRIC SoilGrids platform.

\emph{Use of AI-assisted tools.} In accordance with the IEEE policy on
AI-generated text, the authors disclose that a large-language-model
assistant (Anthropic Claude, via Claude Code) was used in preparing this
work: to assist with the algebra of the QUBO derivations, to identify candidate references and public
data sources, to scaffold the data-acquisition, benchmarking, and analysis
code, and to edit manuscript text. All mathematical
derivations and proofs were independently verified by the authors, all
data provenance and code outputs were validated by the authors, and the
authors take full responsibility for the entire content of this article.
No AI system is an author of this work.

\section*{Data and Code Availability}
All problem-generation, benchmarking, and quantum-execution code, together
with the fetched raw data snapshots (NASA POWER and SoilGrids API
responses) needed to reproduce every figure and table, are available in
the project repository at
\url{https://github.com/mcpeblocker/qubo-for-irrigation-scheduling}.
Quantum hardware results include the IBM job identifiers for independent
retrieval.